\documentclass{emulateapj}

\usepackage{graphicx}
\usepackage{amsmath, amsthm, amssymb,amsfonts}
\usepackage{mathrsfs}
\usepackage{pdfpages}
\usepackage{color}
\usepackage{xspace}

\usepackage{url}

\newcommand{\plot}[1]{\begin{center}\plotone{#1} \end{center}}

\newcommand{\lam}{\lambda}

\renewcommand{\O}{\Omega}
\newcommand{\ep}{\varepsilon}

\newcommand{\al}{\alpha}
\newcommand{\g}{\gamma}
\newcommand{\G}{\Gamma}

\newcommand{\E}{\ep}
\newcommand{\Eiso}{E_\text{iso}}
\newcommand{\gr}{\mbox{$\gamma$-ray}\xspace}

\newcommand{\epg}{\ep_\g}
\newcommand{\epm}{\ep_{\g m}}
\newcommand{\epc}{\ep_{\g c}}
\newcommand{\epn}{\ep_\nu}
\newcommand{\epnc}{\ep_{\nu c}}

\newcommand{\Ginitial}{\G_{i}}
\newcommand{\Gshock}{\G_{s}}
\newcommand{\xie}{\xi_e}
\newcommand{\xib}{\xi_B}
\newcommand{\Lgm}{L_{\g m}}
\newcommand{\xiep}{\xi_e}

\newcommand{\n}{n_0}
\newcommand{\xieN}{\xi_{e,-1}}
\newcommand{\xibN}{\xi_{B,-2}}
\newcommand{\nN}{\n}
\newcommand{\EisoN}{E_{53}}
\newcommand{\TN}{T_1}
\newcommand{\GinitialN}{\left(\frac{\Ginitial}{300}\right)}

\begin{document}

 \title{Ultra High Energy Neutrinos from Gamma-Ray Burst Afterglows using the Swift-UVOT data}
 \author{Guy Nir}
 \affil{Department of Particles and Astrophysics, Weizmann Institute of Science, Rehovot, Israel, 7610001}
 \author{Dafne Guetta}
 \affil{Ort Braude, Karmiel, Israel, 2161002}
 \author{Hagar Landsman}
 \affil{Department of Particles and Astrophysics, Weizmann Institute of Science, Rehovot, Israel, 7610001}
 \and
 \author{Ehud Behar}
 \affil{Department of Physics, Technion, Israel}

\begin{abstract}

  We consider a sample of 107 Gamma Ray Bursts (GRBs) for which early UV emission was measured by {\it Swift}, 
  and extrapolate the photon intensity to lower energies. 
  Protons accelerated in the GRB jet may interact with such photons to produce charged pions and subsequently ultra high energy neutrinos $\ep_\nu\geq 10^{16}$~eV.
  We use simple energy conversion efficiency arguments to predict the maximal neutrino flux expected from each GRB. 
  We estimate the neutrino detection rate at large area radio based neutrino detectors 
  and conclude that the early afterglow neutrino emission is too weak to be detected 
  even by next generation neutrino observatories. 

\end{abstract}

 
\section{Introduction}

  Gamma Ray Bursts (GRBs) are the most powerful explosions in the Universe. 
  The widely used phenomenological interpretation of these cosmological sources is 
  the so called “Fireball (FB) model” \citep{GRB_puzzle_resolved_Piran_2000,GRB_multi_GeV_neutrinos_Meszaros_2000}. 
  In this model the energy carried by the hadrons in a relativistic expanding jet (Fireball) is dissipated 
  internally and distributed between protons, electrons, and the magnetic field in the plasma.
  Part of the bulk kinetic energy is radiated as $\g$-rays (i.e. GRBs) by synchrotron and inverse-Compton radiation of \mbox{(shock-)accelerated} electrons. 
  As the jet sweeps up material it collides with its surrounding medium, which could give rise to Reverse Shocks (RS) and Forward Shocks (FS)~\citep{GRB_reverse_shock_revisited_Gao_Meszaros_2015}. 
  The former may produce an early UV and optical afterglow \citep{GRB_afterglow_neutrino_Waxman_Bahcall_2000}
  while the latter is believed to be responsible for the afterglow emission at longer wavelengths~\citep{GRB_optical_afterglow_Meszaros_1997}.
  The same dissipation mechanism responsible for accelerating electrons that produce the prompt and afterglow photons 
  may also accelerate protons to ultra high energies ($\ep_p\geq 10^{19}$~eV). 
  The interaction of these protons with radiation at the source during the prompt phase~\citep{GRB_fireball_Waxman_Bahcall_1997} and during the afterglow phase~\citep{GRB_afterglow_neutrino_Waxman_Bahcall_2000} could lead to production of charged pions, which subsequently decay to produce neutrinos. 
  
  High energy protons can interact with optical and Ultra-Violet (UV) photons 
  that are radiated by electrons in the reverse shock 
  leading to $\sim 10^{17}$~eV neutrinos via photo-meson interactions~\citep{GRB_afterglow_neutrino_Waxman_Bahcall_2000}. 
  For afterglow emission that peaks at infra-red energies, 
  neutrinos may be produced with energies up to $\sim 10^{19}$~eV. 

  These Ultra High Energy Neutrinos (UHENs) would be delayed 
  with respect to the prompt GRB by the time scale of the RS ($\sim$10-100 s). 
  The same energy conversion efficiency arguments made to assess the neutrino flux 
  from GRB 990123 \citep{GRB_afterglow_neutrino_Waxman_Bahcall_2000} can be used for other GRBs, 
  that have much weaker optical emission, leading to a substantially smaller estimated neutrino flux. 

  The {\it Swift} observatory comprises the $\gamma$-ray Burst Alert Telescope (BAT), 
  which triggers the X-Ray Telescope (XRT), and the UV/Optical Telescope (UVOT)~\citep{GRB_properties_with_swift_Roming_2008}, 
  which provides rapid follow-up observations of GRBs at UV and optical wavelengths.
  Typical time delays from the BAT trigger to first UVOT observations range from 40 to 200~s, 
  making UVOT a good instrument for measuring the early afterglow optical-UV emission. 

  Neutrino astronomy has steadily progressed over the last half century, 
  with successive generations of detectors achieving sensitivity to neutrino fluxes at increasingly higher energies. 
  With each increase in neutrino energy, the required detector increases in size to compensate for the dramatic decrease of the flux.  
  IceCube is a Cherenkov detector~\citep{IceCube_instrument_Halzen_2010} designed specifically to detect neutrinos at GeV-PeV energies. 
  Since May 2011 (\citep{IceCube_completed_2011,IceCube_time_independent_Aartsen_2013}), IceCube has been working with a full capacity of 86 strings, 
  and measured for the first time flux of astrophysical neutrinos. 
  So far no point sources of neutrinos were identified and no correlation with known GRBs were
  found~\citep{IceCube_point_source_search_Kurahashi_2012,IceCube_GRB_search_He_2012,IceCube_GRB_search_results_Whitehorn_2012,IceCube_GRB_prompt_neutrino_search_Aartsen_2015}.
  Antarctic ice allows for an efficient area coverage that makes it possible to construct detectors 
  of order tens to hundreds of km$^2$, and several small-scale pioneering efforts to develop this approach 
  exist~\citep{RICE_results_future_prospects_Kravchenko_2012,AURA_extension_Landsman_2008,ANITA_Gorham_2010}. 
  A modular, radio Cherenkov emission based experiment, the Askaryan Radio Array (ARA) was initiated four years ago. 
  The current stage includes two functioning stations out of the planned 37. 
  The complete detector, ARA37, would cover a hexagonal grid of $\sim 100$~km$^{2}$, 
  and is designed to ultimately accumulate hundreds of cosmogenic neutrinos~\citep{ARA_collaboration_testbed_limits_2014}. 
  The Antarctic Impulsive Transient Antenna (ANITA) experiment, 
  based on a balloon flying over the Antarctic to detect neutrino hits 
  using radio Cherenkov radiation, has already accumulated data in three flights. 
  Neither experiment has yet to detect a high energy neutrino signal. 

  In this work, we exploit the optical and UV data from the {\it Swift}/UVOT to infer the neutrino flux from each GRB, and estimate the probability that these neutrinos would be detected by future large scale observatories.

  In Sec.~2 we introduce the selected GRB sample. 
  In Sec.~3 we describe the model and asses its parameters.
  In Sec.~4 we describe the resulting prediction for neutrinos, and discuss their consequences for the model.

\section{UVOT Sample}
  
  The present UVOT sample includes long GRBs ($2 < T_{90} < 700$~s) detected by {\it Swift} from March 2005 to November 2014. 
  We take only UVOT detections that started less than 200 seconds after the BAT trigger, and UVOT exposures $T_{\text{exp}} \leq 300$~s. 
  We only use GRBs with known redshifts.
  and exclude GRBs for which only upper limits are provided. 
  For each GRB we use the filter effective area and magnitude to calculate the photon count. 
  We calculate the flux by dividing the photon count by the estimated length of the reverse shock, or the total exposure time, whichever is shorter. 
  We use the BAT fluence (in the $15-150$~keV band) as well as the GRB duration, $T_{90}$, the time at which the BAT measured flux drops down to 90\%. 
  All data are taken from the Goddard Space Flight Center website\footnote{Website: \url{http://swift.gsfc.nasa.gov/archive/grb\_table/}}.
 
\begin{figure}
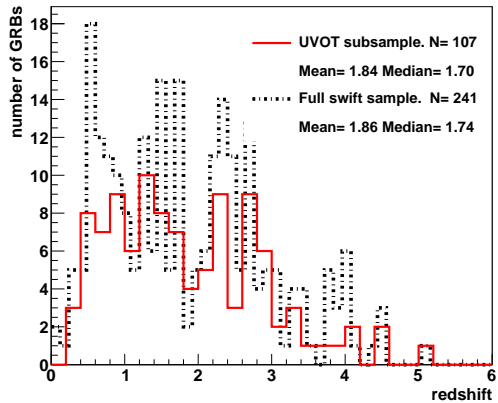

 
    \centering
 
  
      \plot{compare_redshifts}
      \caption{Redshift distribution for the full sample (black line) 
      and for the subset for which exists a UVOT detection (red line). 
      The two sample have a similar distribution, 
      as can be seen by comparing the mean and median values of the two samples. }
      \label{compare_redshifts}
 
    
    \end{figure}

  Our sample includes 107 GRBs (out of $\sim 900$ {\it Swift} bursts).
  The redshift distribution of the present sample is essentially identical to the full {\it Swift} sample.
  The two distributions are plotted in Figure~\ref{compare_redshifts}
  along with the mean/median figures for the full \emph{swift} sample vs.~the chosen subsample. 
  The requirements for early detection and for redshift measurement are due to observational limitations, 
  and do not bias the sample beyond the {\it Swift} field of view and sensitivity limitations. 
 
 
%
  
  \begin{figure}
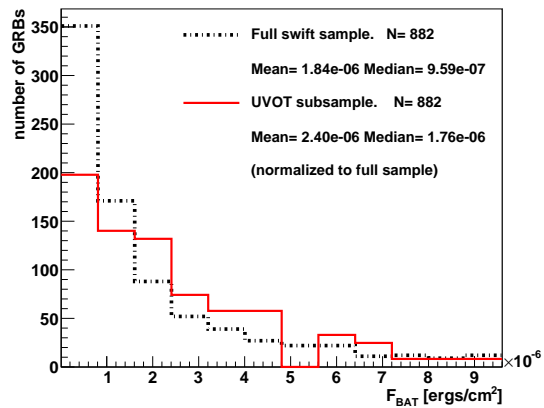

  
      \plot{compare_bat_fluence2}
      \caption{Distribution of BAT fluence for all GRBs detected by \textit{Swift} 
      compared with the present subsample, 
      which is factored up by the number ratio of the two samples.
      The sample is biased by a factor of 0.8 towards brighter events.
      }
      \label{compare_bat_fluence}
 
 
 
  \end{figure}

The mean (median) BAT fluence of the present sample is 75\% (55\%) that of the full sample, see Figure~\ref{compare_bat_fluence}.
Since our neutrino flux estimate scales with the GRB UV luminosity, 
the moderate bias towards high luminosity GRBs in our sample increases the expected mean neutrino flux 
and so should be considered as an upper limit estimate for the neutrino luminosity of the full GRB population. 

\section{Model parameters}
 
  The BAT measured fluence $F_{\text{BAT}}$ can be converted into an isotropic equivalent $\gr$ energy at the source, 
  \begin{equation}\label{eq: distance r squared law}
    E_{\gamma} = \frac{4\pi}{1+z} d_L^2 \ F_{\text{BAT}}
  \end{equation}
  based on the luminosity distance $d_L$ and measured redshift $z$.
  The luminosity distance is calculated using the cosmological parameters~\citep{cosmological_parameters_PDG}: 
  $\O_m=0.3, \O_\Lambda=0.7$, and $H_0= 73.8$ km s$^{-1}$ Mpc$^{-1}$.

  We adopt the hypothesis of the present model that the total $\gr$ energy is equal to the electron energy $E_e$, 
  since in the prompt emission phase the electrons cool much faster than the dynamical timescale. 
  We define $\xiep$ to be the fraction of the total energy $E_e / (E_p + E_e)$ carried by the electrons, 
  where $E_p$ and $E_e$ represent the total energy in protons and in electrons, respectively.
  $E_p$ includes all proton energies from $\E_{p,min}=\Gamma m_p$ up to $\E_{p.max}=10^{22}$~eV.
  The proton flux model is assumed to follow a power law with slope $\al = -2$. 
  We assume all GRBs have the same $\xiep \approx 0.1$ (e.g.~\citep{GRB_energy_budget_Wygoda_2015}) so that 
  the total proton energy is determined directly by the BAT fluence measurement $E_p = 9E_\gamma$. 
  Assuming a single $\xiep$ value is a simplification 
  that yields a sample mean of $\langle E_p \rangle = 10^{53}$~erg
  (for $\E_p\geq10^{19}$~eV), which allows GRBs to be the source of high energy cosmic rays~\citep{GRB_cosmic_ray_origin_Waxman_1995}.

  
  Both the FS and the RS can contribute to the early ($t\leq 200$~s) optical-UV afterglow, 
  it is not clear, however, if the FS can accelerate protons to energies that could yield ultra-high energy neutrinos. 
  In any case, the symplifying assumption that the optical-UV flux is due to the RS gives only an upper limit on the neutrino flux. 
  
\begin{figure}
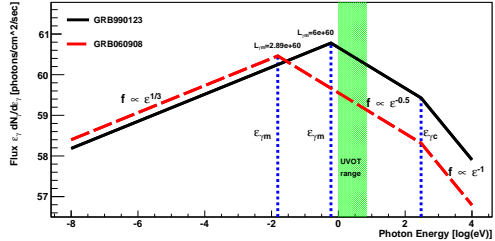

 
    \centering
 
    \plot{plot_example_photon_spectrum}
    \caption{Two example GRB synchrotron spectra. 
    The slow rise, peak energy $\epm$ and cooling break $\epc$ are shown on the plot. 
    The very bright GRB990123 has peak luminosity only $\sim 2$ times 
    stronger than the example GRB050319 from our sample, 
    but has much lower peak energy $\epm$, 
    and thus a much lower UVOT band fluence.  }
 
    \label{plot_example_photon_spectrum}
 
  \end{figure}

  The photon spectrum can be described as a broken power law as expected for synchrotron emission (see Figure~\ref{plot_example_photon_spectrum}).  
  The energy at which this emission peaks is
  \begin{align}\label{eq: peak synchrotron}
    \epm^{ob} = \hbar \Gamma \gamma ^2 \frac{3eB}{2m_e c} = 0.6 \xieN^2 \xibN^{1/2} \nN \GinitialN^2 \text{ eV} 
  \end{align}
  where $\gamma$ is the typical electron Lorentz factor in the plasma, 
  and the general expression is boosted by the jet Lorentz factor $\Gamma$. 

  The typical values of $\xie=0.1 \xieN$ and $\xib=0.01 \xibN$ have been used, 
  as well as the isotropic equivalent energy $\Eiso=10^{53} \EisoN$~erg, the typical RS time $T=10\TN$~s and the ISM density $\n$ in cm$^{-3}$. 
  The Lorentz factor of the unshocked plasma $\Ginitial\sim 300$ is used.

  The photon spectrum follows an approximate power law $dN/d\ep\propto \ep^\al$ with index $\al=-2/3$ up to the peak energy, 
  beyond which the photon spectrum drops as $\al=-1.5$. 
  At a break energy $\epc = 300 \xibN^{-3/2}\n^{-1} \EisoN^{-1/2} \TN^{-1/2}$~eV the spectrum steepens to $\al=-2$,
  as very energetic electrons tend to cool faster than the dynamical time scale.

  The luminosity density at the synchrotron peak produced by a total $N_e$ number of electrons is 
  \begin{align}
    \Lgm  &= \frac{\sqrt{3}}{2}\frac{\Gamma}{2\pi \hbar} \frac{e^3 B}{m_e c^2} N_e \nonumber\\
          &= 6\times 60\, \xibN^{1/2} \EisoN^{5/4} \TN^{-3/4}  \nN^{1/4} \GinitialN^{-1}\text{ s}^{-1}
  \end{align}
  which again is the general expression, boosted by $\Ginitial$. 
  The specific luminosity depends on the total energy of the burst 
  and other model parameters that cannot vary much between GRBs in the sample, 
  so that the luminosity at the peak changes only by factor of a few. 
  The energy at which the flux peaks, however, is treated as a free parameter, 
  and may take very different values for different GRBs measured.
  
\begin{figure}
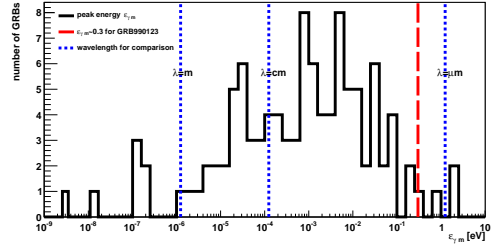

 
    \centering
 
    \plot{hist_peak_photons}
    \caption{Calculated peak syncrotron energy for GRBs in the sample, 
    extrapolated from the UVOT band down to lower energy, 
    assuming a power law of $dN_\g/ d\ep_\g \propto d\ep_\g^{-1.5}$.}
    \label{hist_peak_photons}
 
  \end{figure}

  With UVOT we measure the total energy in the band
  \begin{equation} 
  L_{UV} = \int_{1\text{ eV}}^{7\text{ eV}} L_\g d\epg
  \end{equation}
  so the measured UV luminosity depends on the position of the peak energy (Eq.~\ref{eq: peak synchrotron}), 
  which we find by extrapolating the spectrum from the UVOT band back to lower energies. 
  For a given peak energy, the flux model, extinction corrected, 
  is integrated with the UVOT effective area curve, 
  and compared with the measurement. 
  For each GRB the energy of the peak is adjusted so that the expected and measured fluxes coincide. 
  The calculated $\epm^{ob}$ values for GRBs in our sample are shown in Figure~\ref{hist_peak_photons}. 

\section{Neutrinos from the reverse shock}\label{sec:rs_neutrinos}
 
  Ultra High Energy Neutrinos (UHENu's) may be produced in GRBs 
  through photo-proton interactions that produce charged pions, 
  which in turn decay and emit neutrinos.
  The fraction of proton energy that is transferred to pions 
  depends on the availability of photons at the right energy to produce pions, 
  e.g., through the $\Delta$ resonances~\citep{GRB_afterglow_neutrino_Waxman_Bahcall_2000}. 
  In each interaction a constant fraction ($\sim20\%$) of the proton energy 
  is transferred to the pion, that decays into four particles (three neutrinos and a positron), 
  each getting $\sim 5\%$ of the proton energy.

  The position of the synchrotron peak $\epm$ determines the efficiency 
  for the relevant proton energy~\citep{GRB_afterglow_neutrino_Waxman_Bahcall_2000}, 
  \begin{equation}
   f_\pi(\ep_p) = 0.01 \left(\frac{\Lgm}{10^{60}\text{ s}^{-1}}\right)\left(\frac{\Gshock}{250}\right)^{-5}\TN^{-1} (\ep_{\gamma m}^\text{ob} \ep_{p,20}^{\text{ob}})^{1/2}
  \end{equation}
  for protons at $\ep_p = 10^{20}\ep_{p,20}$~eV, taking a Lorentz factor $\Gshock\sim 250$ for the shocked plasma.
  The efficiency scales linearly with $\Lgm$, 
  but has a very strong dependence on the Lorentz factor. 
  For the typical luminosity density and Lorentz factor of the present sample $\Lgm = 10^{58}$\,s$^{-1}$ and $\Gshock=250$,  
  $f_\pi$ is approximately $10^{-4}$,
  which drives down the pion and neutrino yields considerably. 
 
  Photons above $\epc$ follow a steeper power law, 
  causing a steeper dependence of $f_\pi \propto \ep_p$ for the relevant proton energies. 
  Therefore the neutrino spectrum can be described as a broken power law, 
  following the baseline proton spectrum modulated by the photon density at each energy:
  \begin{equation}
    \frac{dN_\nu}{d\epn} \propto \begin{cases} \epn^{-1} & \epn < \epnc \\ \epn^{-1.5} & \epn > \epnc \end{cases} 
  \end{equation}
  The neutrino break energy is at $\epnc = 2 \times 10^{18}$ eV, corresponding to a photon break energy of $\epc = 300$~eV, and scales inversely with it. 

  \begin{figure}
  
    \centering
  
    \plot{plot_flux_neutrinos}
  
    \caption{Average neutrino quasi-diffuse flux, in thick black line, 
    given by the average of the neutrino fluxes in our sample. 
    The flux is calculated for all flavors of neutrinos and anti-neutrinos combined. 
    Compare the neutrino flux to the ANITA sensitivity from~\citealt{ANITA_Gorham_2010}, in pink full circles, 
    and ARA sensitivity from~\citealt{ARA_collaboration_testbed_limits_2014} in blue empty circles. 
    The diffuse flux from~\citealt{GRB_afterglow_neutrino_Waxman_Bahcall_2000}, in red dotted line, 
    is based on the very bright GRB990123, that is non representative of the sample in this work.}
  
    \label{plot_flux_neutrinos}
  
  \end{figure}
 
  Using these formulae, and based on the observed GRB luminosities and redshift, 
  we calculate for each burst the expected neutrino flux. 
  We estimate the expected quasi-diffuse neutrino flux by multiplying the mean GRB flux 
  by the total number of GRBs all over the sky per year ($1000\text{ yr}^{-1}$), 
  and by dividing by $4\pi$~sr (Figure~\ref{plot_flux_neutrinos}, thick black line).
  When compared with the ANITA sensitivity from \citealt{ANITA_Gorham_2010} (pink, full crosses) 
  and the ARA sensitivity from \citealt{ARA_collaboration_testbed_limits_2014} (blue, empty circles), 
  it is clear that even at high energies, 
  this diffuse neutrino flux is at least four orders of magnitude too weak to be detected by any current or planned detectors.
  Changes to the kinematic parameters, e.g.~the plasma Lorentz factor, 
  make little difference in the overall neutrino flux, 
  and even for very favorable choices the flux of neutrinos is still too low to be  detected. 
 
  In Figure.~\ref{plot_flux_neutrinos} we also plot the diffuse flux 
  estimated in~\citealt{GRB_afterglow_neutrino_Waxman_Bahcall_2000}, as the red dotted line.
  This estimate was based on the assumption that all GRBs are as bright as the single GRB990123 
  that had a peak luminosity of $\Lgm = 10^{60}$\,s$^{-1}$ in the optical band, 
  which is an order of magnitude higher than the luminosities (at equivalent energies) in the UVOT sample. 
  Therefore, it can not represent the sample in this work, or the population of GRBs at large.

\begin{figure}
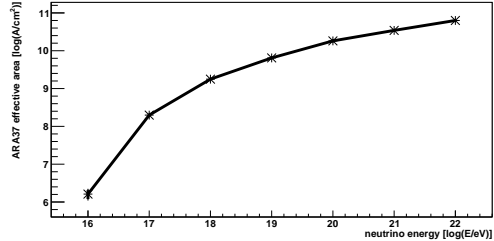

 
    \centering
 
    \plot{plot_ara_aeff}
    \caption{Effective area of ARA37, for all neutrino and anti-neutrino flavors.}
    \label{plot_ara_aeff}
 
  \end{figure}

  Radio frequency high energy neutrino detectors have fairly similar sensitivity to all flavors of neutrinos.
  Hence, neutrino oscillations do not dramatically affect the estimates of detection rates.
  We estimate the expected detection rate for the combined contributions of neutrinos and anti-neutrinos of all flavors. 
  To obtain the number of neutrinos to be measured on Earth, 
  we fold the model neutrino fluence spectrum of a single GRB $dN_\nu / d\epn / dA$ 
  (from Sec.~\ref{sec:rs_neutrinos}) with the energy-dependent ARA37 effective area $A_\text{eff}(\epn)$.
  \begin{equation}\label{eq: number of neutrino detections}
    N_\nu = \int_{\ep_{\nu, min}}^{\ep_{\nu,max}} A_\text{eff}(\epn) \frac{dN_\nu}{d\epn dA} d\epn
  \end{equation}
  The effective area, shown in Figure~\ref{plot_ara_aeff}, 
  is the product of the ARA effective volume~\citep{ARA_NSF_proposal}, 
  the ice density, and the cross section for neutrino absorption by an ice atom nucleon~\citep{cross_section_Connolly}. 
  Below $\epn\sim 10^{16}$~eV the efficiency for radio detection of neutrinos drops very rapidly, 
  while above $\epn\sim 10^{20}$~eV the detector trigger is saturated, 
  and the effective area rises only through the logarithmic increase in cross section.
  The number of neutrinos we expect to detect for an average GRB in our sample is $8.4\times 10^{-7}$. 
  For 1000 GRBs a year (full sky), this implies a total neutrino detection rate of $6.7\times 10^{-5}$\,yr$^{-1}$, 
  thus we conclude that early afterglow neutrinos will not be detectable even in the next generation of neutrino observatories. 

  Looking at extremely bright GRBs both before and during the \emph{swift} era, 
  some bursts show substantial emission at early times that can be attributed to the RS~\citep{GRB_reverse_shock_revisited_Gao_Meszaros_2015}. 
  The peak synchrotron energy and the neutrino flux can be calculated for these bursts 
  using the same calculation made for the entire sample 
  (supplementing data that was unavailable with reasonable estimates).
  For GRB990123 we use the measured magnitude $M=9$ and redshift $z=1.6$~\citep{GRB_990123_ROTSE_observations_Akerlof_1999}
  to recover parameter values similar to those presented in~\citep{GRB_afterglow_neutrino_Waxman_Bahcall_2000}. 
  The number of neutrinos expected in ARA37 from this single GRB would be $N_\nu\sim 2\times 10^{-4}$. 
  If all GRBs had similar parameters, the number of detections per year would be about $N_{GRBs}/4\pi \approx 100$ times this number, 
  still below the detection threshold for ARA.
  
  For GRB080319B, among the brightest GRBs recorded by \emph{swift}, 
  we can only estimate the true magnitude since UVOT had been saturated at $M=13.9$ in the white filter. 
  At this value the number of detections $N_\nu\approx 5\times 10^{-6}$ is not exceptional. 
  Using greater magnitude value for this burst, which is estimated to have peaked at $M\sim 5.3$~\citep{GRB_080319B_naked_eye_Racusin_2008},
  the number of neutrino detections would be  $0.1 \lesssim N_\nu \lesssim 10$, 
  depending critically on the value chosen for the prompt $\xi_e$ and the maximum proton energy. 
  Clearly such a burst is not representative, 
  but had it occurred during the operation period of any large area neutrino detector it may well have been detected. 
  
  The data collected by UVOT for GRB130427A is only available starting at $T=358$~s, making it ineligible for our sample. 
  It is, however, a very bright GRB, and we can assume its brightness at $t\sim 100$~s is similar to the first measurements made. 
  For the magnitude of the first measurement in the V filter $M=12.1$~\citep{GRB_130427A_GCN_2013} we get $N_\nu \approx 5\times 10^{-6}$ neutrino detections. 
  
\begin{figure}
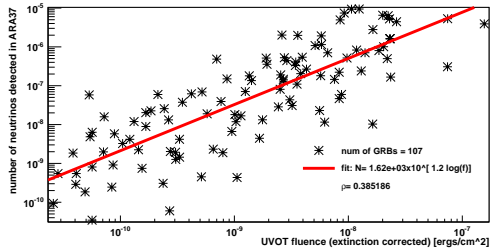

    \centering
    \plot{corr_fluence_nu_hits}
    \caption{Number of neutrino detections expected in ARA37 
    vs.~the fluence measured by UVOT, based on the low peak energy RS model 
    presented in this work. 
    A strong correlation allows an estimate of the number of neutrino detection 
    based on the measured UVOT fluence, for any GRB measured in UVOT. }
    \label{corr_fluence_nu_hits}
  \end{figure}

  In the present sample and within the assumptions of the model, 
  we find that the UVOT fluence is a good predictor of the expected neutrino rate. 
  Although the neutrino flux in the model scales with the total GRB energy, estimated by the BAT $\gr$ fluence, 
  the neutrino flux is strongly modulated by $f_\pi$, 
  which is determined by the intensity of the optical-UV photons available for photon-proton interactions. 
  Figure~\ref{corr_fluence_nu_hits} shows the strong correlation between the fluence measured in UVOT and the total number of expected neutrinos.

\section{Discussion and Conclusions}

  In this work, we calculated the expected ultra high energy neutrino flux from the reverse shock in the early GRB afterglow.
  We use the observed {\it Swift}/UVOT flux as a proxy of the photon energy content in the reverse shock,
  which is the target for the photon-meson production of neutrinos.
  The redshift and BAT fluence distributions of the present UVOT sample are representative of the full {\it Swift} GRB sample, 
  with only a slight bias in favor of brighter events in the BAT fluence distribution. 

  Optical-UV measurements taken within the time frame of the early afterglow ($40\lesssim t \lesssim 200$~seconds) 
  are much lower than anticipated by the RS-optical flash model, 
  suggesting that the peak of the synchrotron emission for most GRBs in the sample is $10^{-6}\lesssim \epm \lesssim 10^{-1}$~eV 
  or $10 \mu m\lesssim \lam_{\g m}\lesssim 1$~m. 
  Measurements in this band, or even in the infra red, taken within the same time window can confirm or rule out this scenario. 
  Alternatively, the RS may occur at a much earlier time (i.e.~$\lesssim 40$~seconds),  
  so that UVOT measurements cannot probe the temporal peak of the emission. 
  In this case the intensity and peak energy of the RS emission may be much higher, 
  as suggested in~\citep{GRB_afterglow_neutrino_Waxman_Bahcall_2000}. 
  Further measurements using faster response detectors 
  or a detection of UHENu's $\lesssim 40$~seconds after the burst could confirm this hypothesis. 
  The low intensity may also indicate the entire paradigm of reverse shock acceleration
  does not explain the early afterglow emission~\citep{GRB_early_afterglow_neutrino_Murase_2007}. 
  Using the UVOT data presented here we cannot rule out any of these possibilities. 
  A detection of an UHENu, 
  as well as measurement of the energy and time delay of such a neutrino 
  would immediately differentiate between these possibilities.   

  The neutrino fluxes obtained based on the low peak energy RS result 
  are approximately four orders of magnitude below the detection sensitivity 
  of present and future high-energy neutrino telescopes.
  This predicted flux from the reverse shock is much lower 
  than that expected from (100~keV) photon meson production in the prompt phase~\citep{GRB_neutrino_BATSE_Guetta_2004,GRB_hadronic_content_Behar_2014}.
  Moreover, two aspects of our analysis may cause an underestimate of the neutrino flux,
  hence, the present estimate provides an upper limit on the neutrino flux, 
  which is, even in the most optimistic case, well below the detection threshold.  

  First, we assume that all of the optical-UV emission is due to the reverse shock,
  and that these photons are in the same region where the protons are accelerated. 
  This implies  that the efficiency for pion production is maximal.
  If part of the UV flux comes from the forward shock, 
  the expected neutrino flux would be even lower than our estimate.
  Furthermore, the present sample is somewhat biased towards high-luminosity GRBs, 
  as we miss the weakest UV sources.
  On average, the present sample is 80\% brighter than the full sample.
  Hence, the average neutrino flux from the full GRB population would be lower by $\sim 0.8$ than estimated here.

  We note that the spectrum of neutrinos due to the prompt emission phase would peak at $\sim 10^{15}$~eV,
  while future radio based neutrino detectors (e.g., ARA) will be more sensitive above $10^{17}$~eV. 
  For a high neutrino flux at these energies, $\sim 10^{19}$~eV protons need to interact 
  with the low-energy (keV) tail of the prompt emission at a sufficiently high rate.
  The fact that prompt neutrinos from GRBs have not yet been detected~\citep{IceCube_GRB_prompt_neutrino_search_Aartsen_2015} 
  together with the low fluxes from the afterglow predicted here,
  imply that ARA may not be optimal for GRB neutrino detection.
  This conclusion is independent of the fact that radio based neutrino observatories are still well suited for detecting cosmogenic neutrinos. 
   
  \acknowledgments
 
  We thank Dave Besson and Markus Ahlers for discussions. This research is supported by a grant from the U.S.~Israel Binational Science Foundation. 
 
  \bibliographystyle{apj}
  \bibliography{Bibliography}

\begin{thebibliography}{}
\expandafter\ifx\csname natexlab\endcsname\relax\def\natexlab#1{#1}\fi

\bibitem[{Aartsen {et~al.}(2013)Aartsen, Abbasi, Abdou, Ackermann, Adams,
  Aguilar, Ahlers, Altmann, Auffenberg, Bai, Baker, Barwick, Baum, Bay, Beatty,
  Bechet, Tjus, Becker, Benabderrahmane, BenZvi, Berghaus, Berley, Bernardini,
  Bernhard, Besson, Binder, Bindig, Bissok, Blaufuss, Blumenthal, Boersma,
  Bohaichuk, Bohm, Bose, Böser, Botner, Brayeur, Bretz, Brown, Bruijn,
  Brunner, Carson, Casey, Casier, Chirkin, Christov, Christy, Clark,
  Clevermann, Coenders, Cohen, Cowen, Silva, Danninger, Daughhetee, Davis, Day,
  Clercq, Ridder, Desiati, de~Vries, de~With, DeYoung, Díaz-Vélez, Dunkman,
  Eagan, Eberhardt, Eisch, Euler, Evenson, Fadiran, Fazely, Fedynitch,
  Feintzeig, Feusels, Filimonov, Finley, Fischer-Wasels, Flis, Franckowiak,
  Frantzen, Fuchs, Gaisser, Gallagher, Gerhardt, Gladstone, Glüsenkamp,
  Goldschmidt, Golup, Gonzalez, Goodman, Góra, Grandmont, Grant, Groß, Ha,
  Ismail, Hallen, Hallgren, Halzen, Hanson, Heereman, Heinen, Helbing,
  Hellauer, Hickford, Hill, Hoffman, Hoffmann, Homeier, Hoshina, Huelsnitz,
  Hulth, Hultqvist, Hussain, Ishihara, Jacobi, Jacobsen, Jagielski, Japaridze,
  Jero, Jlelati, Kaminsky, Kappes, Karg, Karle, Kelley, Kiryluk, Kläs, Klein,
  Köhne, Kohnen, Kolanoski, Köpke, Kopper, Kopper, Koskinen, Kowalski,
  Krasberg, Krings, Kroll, Kunnen, Kurahashi, Kuwabara, Labare, Landsman,
  Larson, Lesiak-Bzdak, Leuermann, Leute, Lünemann, Macías, Madsen, Maggi,
  Maruyama, Mase, Matis, McNally, Meagher, Merck, Meures, Miarecki, Middell,
  Milke, Miller, Mohrmann, Montaruli, Morse, Nahnhauer, Naumann, Niederhausen,
  Nowicki, Nygren, Obertacke, Odrowski, Olivas, Omairat, O'Murchadha, Paul,
  Pepper, de~los Heros, Pfendner, Pieloth, Pinat, Posselt, Price, Przybylski,
  Rädel, Rameez, Rawlins, Redl, Reimann, Resconi, Rhode, Ribordy, Richman,
  Riedel, Rodrigues, Rott, Ruhe, Ruzybayev, Ryckbosch, Saba, Salameh, Sander,
  Santander, Sarkar, Schatto, Scheriau, Schmidt, Schmitz, Schoenen,
  Schöneberg, Schönwald, Schukraft, Schulte, Schulz, Seckel, Sestayo,
  Seunarine, Shanidze, Sheremata, Smith, Soldin, Spiczak, Spiering, Stamatikos,
  Stanev, Stasik, Stezelberger, Stokstad, Stößl, Strahler, Ström, Sullivan,
  Taavola, Taboada, Tamburro, Tepe, Ter-Antonyan, Tešić, Tilav, Toale,
  Toscano, Unger, Usner, Vallecorsa, van Eijndhoven, Overloop, van Santen,
  Vehring, Voge, Vraeghe, Walck, Waldenmaier, Wallraff, Weaver, Wellons, Wendt,
  Westerhoff, Whitehorn, Wiebe, Wiebusch, Williams, Wissing, Wolf, Wood,
  Woschnagg, Xu, Xu, Yanez, Yodh, Yoshida, Zarzhitsky, Ziemann, Zierke, \&
  Zoll}]{IceCube_time_independent_Aartsen_2013}
Aartsen, M.~G., Abbasi, R., Abdou, Y., {et~al.} 2013, The Astrophysical
  Journal, 779, 132

\bibitem[{{Aartsen} {et~al.}(2015){Aartsen}, {Ackermann}, {Adams}, {Aguilar},
  {Ahlers}, {Ahrens}, {Altmann}, {Anderson}, {Arguelles}, {Arlen}, \&
  et~al.}]{IceCube_GRB_prompt_neutrino_search_Aartsen_2015}
{Aartsen}, M.~G., {Ackermann}, M., {Adams}, J., {et~al.} 2015, \apjl, 805, L5

\bibitem[{{Abbasi}(2011)}]{IceCube_completed_2011}
{Abbasi}, R. 2011, ArXiv e-prints, arXiv:1111.5188

\bibitem[{{Connolly} {et~al.}(2011){Connolly}, {Thorne}, \&
  {Waters}}]{cross_section_Connolly}
{Connolly}, A., {Thorne}, R.~S., \& {Waters}, D. 2011, \prd, 83, 113009

\bibitem[{{Gao} \&
  {M{\'e}sz{\'a}ros}(2015)}]{GRB_reverse_shock_revisited_Gao_Meszaros_2015}
{Gao}, H., \& {M{\'e}sz{\'a}ros}, P. 2015, Advances in Astronomy, 2015, 13

\bibitem[{{Gisler} {et~al.}(1999){Gisler}, {Akerlof}, {Balsano}, {Bloch},
  {Casperson}, {Fletcher}, {Gisler}, {Hills}, {Kehoe}, {Lee}, {Marshall},
  {McKay}, {Miller}, {Priedhorsky}, {Szymanski}, \&
  {Wren}}]{GRB_990123_ROTSE_observations_Akerlof_1999}
{Gisler}, G., {Akerlof}, C.~W., {Balsano}, R.~J., {et~al.} 1999, in American
  Institute of Physics Conference Series, Vol. 499, American Institute of
  Physics Conference Series, ed. S.~P. {Brumby}, 82--89

\bibitem[{{Gorham} {et~al.}(2010){Gorham}, {Allison}, {Baughman}, {Beatty},
  {Belov}, {Besson}, {Bevan}, {Binns}, {Chen}, {Chen}, {Clem}, {Connolly},
  {Detrixhe}, {De Marco}, {Dowkontt}, {DuVernois}, {Grashorn}, {Hill},
  {Hoover}, {Huang}, {Israel}, {Javaid}, {Liewer}, {Matsuno}, {Mercurio},
  {Miki}, {Mottram}, {Nam}, {Nichol}, {Palladino}, {Romero-Wolf}, {Ruckman},
  {Saltzberg}, {Seckel}, {Varner}, {Vieregg}, \& {Wang}}]{ANITA_Gorham_2010}
{Gorham}, P.~W., {Allison}, P., {Baughman}, B.~M., {et~al.} 2010, ArXiv
  e-prints, arXiv:1003.2961

\bibitem[{Guetta {et~al.}(2004)Guetta, Hooper, Alvarez-Muniz, Halzen, \&
  Reuveni}]{GRB_neutrino_BATSE_Guetta_2004}
Guetta, D., Hooper, D., Alvarez-Muniz, J., Halzen, F., \& Reuveni, E. 2004,
  Astropart.Phys., 20, 429

\bibitem[{{Halzen} \& {Klein}(2010)}]{IceCube_instrument_Halzen_2010}
{Halzen}, F., \& {Klein}, S.~R. 2010, Review of Scientific Instruments, 81,
  081101

\bibitem[{{He} {et~al.}(2012){He}, {Liu}, {Wang}, {Nagataki}, {Murase}, \&
  {Dai}}]{IceCube_GRB_search_He_2012}
{He}, H.-N., {Liu}, R.-Y., {Wang}, X.-Y., {et~al.} 2012, \apj, 752, 29

\bibitem[{{Karle} {et~al.}(2014){Karle}, {Allison}, {Auffenberg}, {Bard},
  {Beatty}, {Besson}, {Bora}, {Chen}, {Chen}, {Connolly}, {Davies},
  {DuVernois}, {Fox}, {Gorham}, {Hanson}, {Hill}, {Hoffman}, {Hong}, {Hu},
  {Ishihara}, {Kelley}, {Kravchenko}, {Landsman}, {Laundrie}, {Li}, {Liu},
  {Lu}, {Maunu}, {Mase}, {Meures}, {Miki}, {Murchadha}, {Nam}, {Nichol}, {Nir},
  {Pfendner}, {Ratzlaff}, {Richman}, {Rotter}, {Sandstrom}, {Seckel}, {Shultz},
  {Stockham}, {Stockham}, {Sullivan}, {Touart}, {Tu}, {Varner}, {Yoshida}, \&
  {Young}}]{ARA_collaboration_testbed_limits_2014}
{Karle}, A., {Allison}, P., {Auffenberg}, J., {et~al.} 2014, ArXiv e-prints,
  arXiv:1404.5285

\bibitem[{{Karle A.} \& {ARA collaboration}(2013)}]{ARA_NSF_proposal}
{Karle A.}, \& {ARA collaboration}. 2013, {NSF} proposal for {ARA}

\bibitem[{{Kravchenko} {et~al.}(2012){Kravchenko}, {Hussain}, {Seckel},
  {Besson}, {Fensholt}, {Ralston}, {Taylor}, {Ratzlaff}, \&
  {Young}}]{RICE_results_future_prospects_Kravchenko_2012}
{Kravchenko}, I., {Hussain}, S., {Seckel}, D., {et~al.} 2012, \prd, 85, 062004

\bibitem[{{Kurahashi}(2012)}]{IceCube_point_source_search_Kurahashi_2012}
{Kurahashi}, N. 2012, in APS April Meeting Abstracts, C7009

\bibitem[{{Lahav} \& {Liddle}(2014)}]{cosmological_parameters_PDG}
{Lahav}, O., \& {Liddle}, A.~R. 2014, ArXiv e-prints, arXiv:1401.1389

\bibitem[{{Landsman} {et~al.}(2009){Landsman}, {Ruckman}, {Varner}, \& {IceCube
  Collaboration}}]{AURA_extension_Landsman_2008}
{Landsman}, H., {Ruckman}, L., {Varner}, G.~S., \& {IceCube Collaboration}.
  2009, Nuclear Instruments and Methods in Physics Research A, 604, 70

\bibitem[{{Maselli} {et~al.}(2013){Maselli}, {Beardmore}, {Lien}, {Mangano},
  {Mountford}, {Page}, {Palmer}, \& {Siegel}}]{GRB_130427A_GCN_2013}
{Maselli}, A., {Beardmore}, A.~P., {Lien}, A.~Y., {et~al.} 2013, GRB
  Coordinates Network, 14448, 1

\bibitem[{Meszaros \& Rees(2000)}]{GRB_multi_GeV_neutrinos_Meszaros_2000}
Meszaros, P., \& Rees, M. 2000, Astrophys.J., 541, L5

\bibitem[{{M{\'e}sz{\'a}ros} \&
  {Rees}(1997)}]{GRB_optical_afterglow_Meszaros_1997}
{M{\'e}sz{\'a}ros}, P., \& {Rees}, M.~J. 1997, \apj, 476, 232

\bibitem[{Murase(2007)}]{GRB_early_afterglow_neutrino_Murase_2007}
Murase, K. 2007, Phys.Rev., D76, 123001

\bibitem[{{Piran}(2000)}]{GRB_puzzle_resolved_Piran_2000}
{Piran}, T. 2000, \physrep, 333, 529

\bibitem[{{Racusin} {et~al.}(2008){Racusin}, {Karpov}, {Sokolowski}, {Granot},
  {Wu}, {Pal'Shin}, {Covino}, {van der Horst}, {Oates}, {Schady}, {Smith},
  {Cummings}, {Starling}, {Piotrowski}, {Zhang}, {Evans}, {Holland}, {Malek},
  {Page}, {Vetere}, {Margutti}, {Guidorzi}, {Kamble}, {Curran}, {Beardmore},
  {Kouveliotou}, {Mankiewicz}, {Melandri}, {O'Brien}, {Page}, {Piran},
  {Tanvir}, {Wrochna}, {Aptekar}, {Barthelmy}, {Bartolini}, {Beskin}, {Bondar},
  {Bremer}, {Campana}, {Castro-Tirado}, {Cucchiara}, {Cwiok}, {D'Avanzo},
  {D'Elia}, {Della Valle}, {de Ugarte Postigo}, {Dominik}, {Falcone}, {Fiore},
  {Fox}, {Frederiks}, {Fruchter}, {Fugazza}, {Garrett}, {Gehrels},
  {Golenetskii}, {Gomboc}, {Gorosabel}, {Greco}, {Guarnieri}, {Immler},
  {Jelinek}, {Kasprowicz}, {La Parola}, {Levan}, {Mangano}, {Mazets},
  {Molinari}, {Moretti}, {Nawrocki}, {Oleynik}, {Osborne}, {Pagani}, {Pandey},
  {Paragi}, {Perri}, {Piccioni}, {Ramirez-Ruiz}, {Roming}, {Steele}, {Strom},
  {Testa}, {Tosti}, {Ulanov}, {Wiersema}, {Wijers}, {Winters}, {Zarnecki},
  {Zerbi}, {M{\'e}sz{\'a}ros}, {Chincarini}, \&
  {Burrows}}]{GRB_080319B_naked_eye_Racusin_2008}
{Racusin}, J.~L., {Karpov}, S.~V., {Sokolowski}, M., {et~al.} 2008, \nat, 455,
  183

\bibitem[{{Roming} {et~al.}(2008){Roming}, {Koch}, {Oates}, {Porterfield}, \&
  {vander Berk}}]{GRB_properties_with_swift_Roming_2008}
{Roming}, P.~W.~A., {Koch}, T.~S., {Oates}, S.~R., {Porterfield}, B.~L., \&
  {vander Berk}, D.~E. 2008, in American Institute of Physics Conference
  Series, Vol. 1065, American Institute of Physics Conference Series, ed. Y.-F.
  {Huang}, Z.-G. {Dai}, \& B.~{Zhang}, 81--84

\bibitem[{{Waxman}(1995)}]{GRB_cosmic_ray_origin_Waxman_1995}
{Waxman}, E. 1995, \apjl, 452, L1

\bibitem[{{Waxman} \& {Bahcall}(1997)}]{GRB_fireball_Waxman_Bahcall_1997}
{Waxman}, E., \& {Bahcall}, J. 1997, Physical Review Letters, 78, 2292

\bibitem[{Waxman \& Bahcall(2000)}]{GRB_afterglow_neutrino_Waxman_Bahcall_2000}
Waxman, E., \& Bahcall, J. 2000, The Astrophysical Journal, 541

\bibitem[{{Whitehorn}(2012)}]{IceCube_GRB_search_results_Whitehorn_2012}
{Whitehorn}, N. 2012, Journal of Physics Conference Series, 375, 052033

\bibitem[{{Wygoda} {et~al.}(2015){Wygoda}, {Guetta}, {Mandich}, \&
  {Waxman}}]{GRB_energy_budget_Wygoda_2015}
{Wygoda}, N., {Guetta}, D., {Mandich}, M.-A., \& {Waxman}, E. 2015, ArXiv
  e-prints, arXiv:1504.01056

\bibitem[{{Yacobi} {et~al.}(2014){Yacobi}, {Guetta}, \&
  {Behar}}]{GRB_hadronic_content_Behar_2014}
{Yacobi}, L., {Guetta}, D., \& {Behar}, E. 2014, \apj, 793, 48

\end{thebibliography}

\end{document}